\documentstyle[11pt,newpasp,twoside]{article}
\def \kms       {km~s$^{-1}~$}
\def \kmsMpc    {km~s$^{-1}$~Mpc$^{-1}~$}

\def\edcomment#1{\iffalse\marginpar{\raggedright\sl#1\/}\else\relax\fi}
\reversemarginpar

\begin{document}
\title{Galaxy Collisions and Star Formation}
\author{Susan A. Lamb}
\affil{University of Illinois, Center for Theoretical Astrophysics,
Departments of Physics and of Astronomy, 1110 W. Green Street, Urbana, IL
61801. Email:slamb@astro.uiuc.edu}

\author{Nathan C. Hearn}
\affil{University of Illinois, Center for Theoretical Astrophysics,
Department of Physics, 1110 W. Green Street, Urbana, IL 61801, USA}

\begin{abstract}

We present a brief overview of some recent observations of colliding
galaxies and relevant numerical simulations. These are compared,
and details of the locations and history of collision induced star
formation are explored, with possible application to star formation 
at earlier epochs.

\end{abstract}


\section{Introduction}

Galaxy collisions may be the predominant cause of star
formation in galaxies  over the course of their lifetimes. In
the local universe, strong collisions between comparable mass
galaxies are rare, and star formation in gaseous disks
usually takes place in relatively quiescent circumstances.
However, it  is thought that there was an epoch in the past
when both strong and glancing collisions between comparable
mass galaxies were much more common. These interactions would
have produced strong tidal torques and density waves, as
observed in the rare, strongly interacting disk galaxies of
the local universe.  Observations of recently impacted disk
galaxies usually show evidence of enhanced rates of recent
and current star formation (starbursts). Our
three-dimensional numerical simulations of collisions between
comparable mass galaxies demonstrate that the gas volume
density in the disk can be increased by significant factors,
both in the nucleus and in pronounced features that form well
away from the galactic center.  These collision-induced
density enhancements can give rise to ring galaxies, strong,
one-armed spirals, and grand-design spirals.  The very high
density regions produced experience strong shocks in many
circumstances, which may play an important role in
determining the subsequent amount and exact location of star
formation.  We present comparisons between multi-wavelength
observations of star forming, impacted galaxies and our 3-D
numerical simulations of galaxy collisions involving disk
galaxies. We show that the timescale for massive star
formation can be very short, and that the resulting
morphology and velocity structure in the disk can be
understood for several well-observed systems.  

The collision and subsequent merger of clumps of stars and
gas to form the current galaxies is a central tenet of the
theory of hierarchical galaxy formation. (This is reviewed
elsewhere in this volume.) However, the details of the
interactions are somewhat unclear, as yet. Do the "dark
matter" halos merge at an early stage, while luminous clumps
of material have a large relative velocity within these
halos, forming tight, high velocity galaxy groups? Are the
currently observed compact groups (see Sulentic 2000) the
remnants of such a population? The amount and location of
star formation in this interacting material, as in current
galaxies, will be effected by, among other things, the
relative velocities and morphologies of the colliding
galaxies, the amount of molecular gas that exists, and the
background gravitational potential. The effects of these
parameters can be explored for systems in the nearby universe
and the results applied judiciously to studies of star
formation at earlier times. 

One area of exploration that has been given considerable
observational attention is that of the relationship between
interactions, star formation, and the presence of molecular
gas. Recently, this has been investigated for an
optically-selected sample of interacting galaxies (see
Bushouse et al. 2000, and references within).  The choice of
sample is important because many previous studies were of
infrared-bright selected galaxies which might give somewhat 
different results. (There are numerous references on this 
subject, for example Young et al. 1996.) The Bushouse sample
spans a large range of interaction strengths and star
formation rates, as well as infrared properties. Several of
the conclusions drawn were consistent with those found from
the IR-selected samples, namely, that 1) interacting
galaxies are, on average, marginally richer in molecular gas
than a comparison sample of isolated spiral galaxies
(assuming the standard conversion factor of CO to H$_2$), 2)
the interacting galaxies also have a mean infrared
luminosity to H$_2$ mass ratio that is a factor of ~1.3
higher than the isolated galaxies, and 3) there is a strong
correlation between relative H$_2$ content and the star
formation rate, indicating that interaction-induced star
formation activity is highly dependent upon the available gas
supply. However, other results could only be found in a
study of this type of sample. These included, that some
galaxies have moderate amounts of H$_2$ but much lower than
normal star formation rates and IR emission. (This implies
that molecular gas is a necessary, but not sufficient,
prerequisite for star formation in these systems.) An
apparent strong correlation between interaction strength and
both star formation rate and relative H$_2$ content is found
in these studies. However, this correlation could be
completely removed by invoking non-standard CO-to-H$_2$
conversion factors. Thus there is no strong evidence that
interactions initiate conversion of HI to H$_2$ gas, only
that the star formation rate is usually higher when more
H$_2$ is present.

\section{Detailed exploration of nearby systems}

Observations of recently impacted objects in the local
universe, when compared  with numerical simulations of galaxy
collision, can be used to investigate some important aspects
of the induced star formation. In this section we review the
results from two studies involving comparisons between
multiwavelength observations of star-forming impacted disk
galaxies and numerical simulations of colliding galaxies.
The systems Arp 118 and Arp 119 both
contain a strongly perturbed disk galaxy and an identifiable
intruding companion which collided recently, that is, within
the last few tens of millions of years. In the studies of
these systems, it has been shown that fully
dynamical, 3D numerical simulations of an elliptical galaxy
colliding with a disk galaxy can reproduce the morphology of
the disturbed disk galaxy and the general shape of the
elliptical. The results of the simulations
are consistent with the approximate relative positions of the
two galaxies within each pair, and with the velocity gradients
across the disks. The ability to simulate the gross behaviors
of both the stars and gas in specific colliding systems leads
to the possibility of investigating the timing and amount of
massive star formation following the collision, of
identifying the size of some of the large-scale physical
parameters of the gas involved in the new star formation, and
of following the resulting flows of gas in the impacted disk.

All of the 'best-fit' models to the colliding pairs mentioned
above were  chosen from a grid of simulations presented in
Gerber (1993), Gerber, Lamb, \& Balsara (1996), and Lamb,
Hearn, \& Gerber (2000). These simulations explore the
results of face-on collisions (ones parallel to the spin-axis
of the disk) between an elliptical galaxy and a disk galaxy,
and are comprised of combined N-body/hydrodynamical
simulations. The model of the rotating disk galaxy consists
of gas, stars, and dark matter, while that of the spherical
galaxy is composed of stars and dark matter only. Mass ratios
from 1:1 to 10:1 have been explored, with impact parameters
ranging from zero (head-on) to slightly larger than the disk
radius. The gas dynamics is treated using the smoothed
particle hydrodynamics (SPH) method, and the gravitational
forces are calculated using standard particle-mesh (PM)
techniques. Both the n-body (stellar and dark matter) and SPH
(gas) particles contribute to the gravitational potential. 

The numerical simulations can be physically scaled to
individual galaxy systems if the masses and sizes of the
galaxies involved are known. Thus, an estimate of the
distance, as well as observations at a variety of wavelengths,
must be available if a model fitting is to be attempted for
a particular system. The choice of real systems to compare
with simulations is mostly dictated by the existence of
observational data sets for comparable mass galaxies in
pairs.  Further selection is based upon the general
morphology of the disk galaxy, where indications that the
impact was roughly parallel to the spin axis of the disk is
sought. Such an impact produces several unique morphological
features, such as full or partial rings. These impact-induced
morphologies are explored and discussed in Gerber \& Lamb
(1994). Once the present-day match is found between a real
system and a model taken from a simulation, observed regions
of recent, current, and potential star formation can be
matched with regions of high gas density and shocks during the
course of the simulation. Some understanding of the history
of, and the conditions influencing, star formation can then
be obtained from the simulation.

Arp 118 is comprised of two galaxies, an elliptical
(NGC~1143) and a disturbed disk galaxy (NGC 1144) 
containing a very extended starburst ring and connected arc.
These two galaxies have a separation of 42 kpc on the sky
(assuming $H_{0}=75$ \kmsMpc), and a relative velocity
between their nuclei of 300 \kms along the line of sight.
NGC~1143 is a luminous infrared galaxy with a total
luminosity of approximately $3\times10^{11}~$L$_{\odot}$ and
L$_{IR}=2.5\times10^{11}~$L$_{\odot}$ (Appleton \&
Struck-Marcel 1987), a very strong CO(1-0) emission
(L$_{CO}=10^{10}~$L$_{\odot}$, Gao et al. 1997), and an
extreme velocity gradient across the ring of 1100 \kms (see
Hippelein 1989a, b). Recent observations of the neutral and
ionized gas in this system are presented in Bransford et al.
(1999).  The nucleus is displaced from the center of the ring
structure and is classified as Seyfert 2 (Hippelein 1989a).

Lamb, Hearn, \& Gao (1998) showed that the disturbed
morphology, the extreme velocity gradients, and the
morphological relationships between the emitting regions as
observed in radio, H$\alpha$, and CO, can be explained as
resulting from a  collision between a rotating disk galaxy
and a somewhat less massive elliptical which passed through
the disk approximately 42 Myr ago at a slight angle to the
perpendicular to the disk. The detailed modeling does not
support a prior suggestion (see Gao 1996, and references
within) that a merger was needed to explain the morphology
and large velocity gradients of NGC 1144. The overall
dynamics and morphology of the Arp 118 system are well
represented by a numerical model of an off-center collision
in which the disk galaxy has a mass four times that of the
intruding elliptical. In particular, the most notable
large-scale feature of the system, the extensive arm of
compressed gas edged with a dark dust lane, is very well
matched by the model.

The sequence of the various star formation episodes in
NGC~1144 was determined by matching the models at three
different stages of the simulation with observations in three
wavelength bands mentioned above, each of which corresponds
to a well-defined epoch of star formation. The radio  and
H$\alpha$ emission of a starbursting galaxy are an indication
of the star formation activity that occurred at earlier
times, the H$\alpha$ being from the more recently formed
stars. The CO observations provide information about those
regions of the gas that are currently dense and likely to
form stars shortly. The dimensions of the simulation were
scaled to the physical system using the observed length
scales and mass of the spiral galaxy as given by Gao et al. 
(1997) and Hippelein (1998a). These scaled units
were then used to determine the times at which the collision
and subsequent star formation episodes took place.

This study demonstrates for the first time the connection
between the spatial and temporal progression of star
formation, and the changing locations of the very dense
regions in the gas  of a massive disk galaxy in
the aftermath of its collision with a massive elliptical. The
collision generated a strong, non-linear density wave in the
stars and gas in the disk of NGC~1144, causing the gas to
became clumped on a large scale.  This wave produced a series
of superstarclusters along arcs and  rings that emanate from
the central point of impact.  The locations of these star
forming regions match those of the  regions of increased gas
density predicted in the time sequence of models.

It is thought that stars, particularly massive stars, are
formed in the cores of giant molecular clouds in the highest
density regions. Both multiwavelength observations and the
numerical models indicate that strong shocks in the gas,
together with large increases in the gas volume density, are
associated with star formation over volumes of 1 kpc$^3$. The
observed morphology of the regions of dense  gas and the
clustering observed in the newly formed, H$\alpha$
emitting stars suggest that the observed clumping of the
young stars results from a clumping of the densest gas on the
same scale. The work by Marston \& Appleton (1995) and
Appleton \& Marston (1997) also provides evidence that the
clumping observed in the optical images of collisionally
produced ring galaxies is not due to patchy dust obscuration,
because the same clustering is also observed in the near
infrared. Gas clumping on this same scale is found in the
numerical simulations, suggesting that there is a global
explanation for the observed morphology of the dense gas and
the resulting giant stellar formations in these systems.

The Arp 119 system (also known as CPG 29) has also been
studied. The southern member (Arp 119S) of the pair has a
strongly disturbed, asymmetric form, while the northern
member, Arp 119N, appears to be a normal elliptical, although
very devoid of gas (Marziani, et al, 1994). Arp 119S exhibits
some remarkable features, including: a bright, circular ring
surrounding the nucleus; a luminous arc north of the nucleus
that extends to the west; a long arc of knots along the
southern edge; and a ÒveilÓ of low-luminosity material in the
east. This galaxy nucleus is classified as a LINER.
Observations and model fitting of this system are summarized
in another paper in this volume (see Hearn, Lamb, Gruendl, \&
Gerber 2000). A longer paper describing new infrared
observations and a simulation fit to these and other
observations is in preparation (Hearn \& Lamb 2000).

The displacement of the nucleus in Arp 119S from the center
of the ring indicates a face-on collision with a significant
impact parameter (about 25 percent of the disk's radius), and
the structure of the luminous arc in the northern part of Arp
119 indicated that the elliptical and disk galaxies,
including their dark matter halos, are of comparable mass.
The analysis of the simulations and the observations suggests
that the collision between Arp 119N and Arp 119S took place
about 71 million years ago. Multiwavelength image, see, for example, Marziani
et al. 1994, reveal locations of past and present starburst events in Arp 119S. 
These regions have been correlated in space and time with
regions of enhanced density and strong  shocks in the
simulated gas disk.  It was found that at least two distinct
star formation episodes occurred as a result of the
collision. One is currently ongoing, and the other occurred
approximately 24 Myr after the impact. The first burst
occurred when the expanding, rotating structure of dense gas
produced by the collision had formed a ring around the inner
third of the disk. This ring of gas produced the bright ring
of stars around the nucleus that is now observed in the
B-band image. This high density region in the gas has now
spread and opened up, and reaches to the edge of the visible
disk. It shows up as a northern arc of molecular gas that
leads to a bright arc of B-band and H$\alpha$ emission. The
long arc of star-forming knots at the southern edge of Arp
119S is coincident with high-density knots in the gas,
forming a southern arc in the simulation. The gas density is
relatively low in the eastern half of the simulation,
matching the region of low luminosity in the optical
image. The study of the colliding pair Arp 119 indicates that
the star formation triggered by a collision is not continuous
in time or space through the impacted disk, but rather occurs
in distinct episodes.

\section{Conclusions}

The studies described above demonstrate that a careful
comparison between high resolution observations and detailed
models can yield   insight into the sequence of star
formation that takes place  in a gas-rich galaxy after a
major collision. The relative velocities  between the two
galaxies can be high, as illustrated by the Arp 119 system
and the observations of compact groups (Sulentic 2000), and
the first encounter can produce considerable morphological
change and trigger star formation. The time for a merger of
the members of the pair or group may be very large, but the
star formation can be triggered early, and perhaps often, by
collisions. 

The intensity and location of the starburst at any particular
epoch will depend upon the speed with which density waves are
propagating through the expanding disk. Such quantities can
now be predicted quite accurately from current models of
colliding galaxies. Thus star formation on the scale of
several hundred parsecs, as it occurs in these systems at the
current epoch, can now be investigated more thoroughly. The
rate of galaxy collisions in the past was larger than it is
today, so a considerable portion of the star formation that
took place in disk galaxies at earlier epochs was likely
triggered by galaxy collisions. We expect, therefore, that
studies like the ones summarized here will help in
understanding this earlier star formation and its current
consequences.

A longer overview of studies of this type, accompanied by a
video illustrating some of the results, is given in Lamb,
Hearn, \& Gerber (2000).

\end{document}